\documentclass[12pt]{article}

\usepackage{amsfonts,amssymb,amsmath,enumerate}
\usepackage{color}
\usepackage{theorem,cite}
\usepackage{indentfirst}
\usepackage{textcomp}

\newtheorem{thm}{Theorem}[section]
\newtheorem{prop}[thm]{Proposition}
\newtheorem{lem}[thm]{Lemma}

\newtheorem{defi}[thm]{Definition}
\newcommand{\prf}{\textsl{Proof:}\ }
\newcommand{\finprf}{\null \hfill {\rule{5pt}{5pt}}\\[2.1ex]\indent}

\theorembodyfont{\rmfamily}

\newcommand{\bea}{\begin{eqnarray}}
\newcommand{\eea}{\end{eqnarray}}
\newcommand{\shit}{\begin{align}}
\newcommand{\mdr}{\end{align}}
\newcommand{\beano}{\begin{eqnarray*}}
\newcommand{\eeano}{\end{eqnarray*}}
\newcommand{\beq}{\begin{equation}}
\newcommand{\eeq}{\end{equation}}

\newcommand{\mb}[1]{\hspace{2.1ex}\mbox{#1}\hspace{2.1ex}}
\numberwithin{equation}{section}

\def\fT{{\mathfrak T}}

\def\ffi{{\mathfrak i}}

\def\fm{{\mathfrak m}}
\def\fn{{\mathfrak n}}

\def\fs{{\mathfrak s}}

    \def\cB{{\cal B}}    
        
    \def\cH{{\cal H}}

\def\cS{{\cal S}}    \def\cT{{\cal T}}    \def\cU{{\cal U}}

\newcommand{\CC}{{\mathbb C}}

\newcommand{\II}{{\mathbb I}}

\begin{document}
 
\begin{center}

 {\LARGE  {\sffamily A new braid-like algebra for  Baxterisation} }\\[1cm]

\vspace{10mm}
  
{\Large 
 N. Crampe$^{a}$\footnote{nicolas.crampe@univ-montp2.fr}, 
 L. Frappat$^b$\footnote{luc.frappat@lapth.cnrs.fr},
 E. Ragoucy$^{b}$\footnote{eric.ragoucy@lapth.cnrs.fr}
 and M. Vanicat$^{b}$\footnote{matthieu.vanicat@lapth.cnrs.fr}}\\[.41cm] 
{\large $^a$ Laboratoire Charles Coulomb (L2C), UMR 5221 CNRS-Universit\'e de Montpellier,\\[.242cm]
Montpellier, F-France.}
\\[.42cm]
{\large $^{b}$ Laboratoire de Physique Th{\'e}orique LAPTh,
 CNRS--Universit{\'e} de Savoie Mont Blanc.\\[.242cm]
 9 chemin de Bellevue, BP 110, F-74941  Annecy-le-Vieux Cedex, 
France. }
\end{center}
\vfill

\begin{abstract}
We introduce a new Baxterisation for $R$-matrices that depend separately on two spectral parameters. 
The Baxterisation is based on a new algebra, close to but different from the braid group. 
This  allows us to recover the $R$-matrix of the multi-species generalization of the totally asymmetric simple exclusion process
with different hopping rates.
\end{abstract}

\vfill\vfill
\rightline{LAPTH-051/15}
\rightline{September 2015}

\newpage
\pagestyle{plain}
\section*{Introduction} 

The concept of R-matrix is nowadays a basic ingredient in the study of quantum integrable systems.
More specifically, the R-matrices depending on spectral parameters constitute the building blocks for constructing the monodromy matrix and then the transfer matrix for such systems. 
It is known that when the R-matrix satisfies the celebrated Yang--Baxter equation (YBE), the transfer matrices commute for different values of the spectral parameter. 
The transfer matrix therefore encodes the conserved commuting quantities of the system, and among them
the Hamiltonian, usually defined as the logarithmic derivative of the transfer matrix at a specific
value of the spectral parameter (in practice the one for which the R-matrix is regular).
Thus, the determination of the possible solutions of the YBE with spectral parameter is of great importance.
Unfortunately, this is an (almost) hopeless task given the functional coupled equations to be solved.
Nevertheless, large classes of R-matrices with spectral parameters are known (e.g. rational or trigonometric, based on simple Lie algebras), see e.g. \cite{KuReSk,KuSkl,J86}.

A particularly interesting technique was proposed by V.F.R. Jones \cite{Jones} in a framework of knot theory, which is known as Baxterisation. It allows one to obtain solutions of the YBE with spectral parameter from representations of the braid group or quotients thereof. 
Important cases are the ones of the Hecke algebra, the Temperley--Lieb algebra or the Birman--Murakami--Wenzl algebra \cite{Jim86,Isaev}. Since then, many authors tried to generalise or produce other suitable formulae that may lead to solutions of the YBE, see e.g. \cite{CGX,ZGB,YQL,BM2000,ACDM,KMN,FRR13}.

Here, we introduce a new braid-like algebra ${\cal S}_n$ (see definition \ref{def:Sn}). We produce, as the main result of the paper (see theorem \ref{th:main}), a new Baxterisation formula that leads to R-matrices depending genuinely on two spectral parameters. The obtained R-matrices satisfy the usual properties of unitarity, regularity and locality. 
Morevover, we show that the matrix representations of ${\cal S}_n$ are determined only by the defining relations of ${\cal S}_3$. A classification thereof in terms of $4 \times 4$ matrices is given together with the expressions of the corresponding R-matrices. 
Following the mainstream of quantum integrable systems, we also get the corresponding Hamiltonians.
In the general case, a particular $m \times m$ representation is exhibited, that appears to be linked to some generalisations of the multi-species 
Totally Asymmetric Simple Exclusion Process (TASEP).

\section{A new braid-like algebra}

\subsection{Definition of the algebra} 

We introduced here the different algebras as well as some of their properties used in the rest of the paper.
\begin{defi}\label{def:Sn}
For any integer $n\geq1$, $\mathcal{S}_n$ is the unital associative algebra over $\CC$ with generators $\sigma_{1},\dots,\sigma_{n-1}$ and subject to
the relations
 \begin{eqnarray}
&&  [\sigma_{i+1}\,\sigma_{i}\,,\,\sigma_{i}+\sigma_{i+1}]=0\,,\qquad i=1,...,n-2 \label{eq-Sn-1}\\
&&  [\sigma_{i}\,,\,\sigma_{j}] = 0\,,\quad |i-j|>1 \label{eq-Sn-2}
 \end{eqnarray}
 where $[\,.\,,\,.\,]$ stands for the commutator. 
\end{defi}
Let us stress that the algebra $\cS_n$ is defined without the inverse generators $\sigma_i^{-1}$:
there are interesting realizations of this algebra where represented generators are non-invertible (see e.g. proposition \ref{prop:S} below). For $n=1$, one has $\cS_1 \simeq \CC$.

Relation \eqref{eq-Sn-1} can be written equivalently as
\begin{equation}
 \sigma_{i}\,\sigma_{i+1}\, \sigma_{i}-\sigma_{i+1}\,\sigma_{i}\, \sigma_{i+1}=\sigma_{i+1}\,\sigma_{i}^2-\sigma_{i+1}^2\,\sigma_{i}\;.
\end{equation}
Note that imposing the L.H.S. of the previous relation to be zero, one recovers the defining relations of the braid group $\cB_n$ (however without the inverse generators). This justifies the terminology we used for the algebra $\cS_n$ as a braid-like algebra.
Let us also mention that $\cS_n$, like $\cB_n$, is infinite dimensional.

\begin{prop}\label{pro:iso2}
 The M\"obius map 
\begin{eqnarray}
 \fm_{\alpha,\beta,\gamma} : \mathcal{S}_n&\rightarrow &\mathcal{S}_n\nonumber\\
 \sigma_{i} &\mapsto &(\alpha\ + \beta \sigma_i)\ (1+\gamma \sigma_i)^{-1}\label{eq:iso2}
\end{eqnarray}
with $\alpha, \beta, \gamma \in \CC$ is an algebra homomorphism.
The inverse in \eqref{eq:iso2} has to be understood as the following formal series:
\begin{equation}
 (1+\gamma \sigma_i)^{-1}=\sum_{k=0}^\infty \left(-{\gamma}\ \sigma_i\right)^k\;.
 \label{def:inverse}
\end{equation}
\end{prop}
 \prf We divide the proof into two parts, depending on whether $\gamma$ is null or not. 
 
When $\gamma\neq 0$, the M\"obius transformation is the composition of the following maps 
\begin{equation}
\fm_{\alpha,\beta,\gamma}=\fs_{\beta/\gamma,\alpha-\beta/\gamma}\   \circ\ \ffi\ \circ\ \fs_{1,\gamma} 
\end{equation}
where $\fs_{\alpha,\beta}(\sigma)=\alpha+\beta\sigma$ and $\ffi(\sigma)=\sigma^{-1}$. It is straightforward to show that $\fs_{\alpha,\beta}$ is an homomorphism. To prove that $\ffi$ is also an homomorphism, we start from invertible
elements $\sigma_i$ verifying \eqref{eq-Sn-1} and multiply this relation on the right by $\sigma_i^{-1}\sigma_{i+1}^{-1}\sigma_i^{-1}$ and on the left by 
$\sigma_{i+1}^{-1}\sigma_{i}^{-1}\sigma_{i+1}^{-1}$. This shows that $\sigma_i^{-1}$ verify also \eqref{eq-Sn-1}. Note that 
$\fs_{1,\gamma}(\sigma)$ is an invertible element according to the expansion \eqref{def:inverse}.
Then $\fm_{\alpha,\beta,\gamma}$ is 
an homomorphism since it is a composition of homomorphisms.

For $\gamma=0$, one gets
$\fm_{\alpha,\beta,0}=\fs_{\alpha,\beta}$
which is an homomorphism.
\finprf

We also introduce another algebra $\mathcal{T}_n$, defined as follows
\begin{defi}
For any integer $n\geq1$, $\mathcal{T}_n$ is the unital associative algebra over $\CC$ with generators $\tau_{1},\dots,\tau_{n-1}$ and subject to
the relations
 \begin{eqnarray}
&&  [\tau_{i}\, \tau_{i+1}\,,\,\tau_{i}+\tau_{i+1}]=0\,,\qquad i=1,...,n-2 \label{eq-Snp-1}\\
&&  [\tau_{i}\,,\,\tau_{j}] = 0\,,\quad |i-j|>1. \label{eq-Snp-2}
 \end{eqnarray}
\end{defi}
This algebra is closely related to $\cS_n$ as stated in the following proposition:
\begin{prop}\label{pro:iso}
 $$
\mbox{The map }\quad
\phi :\quad \begin{cases} \mathcal{S}_n&\rightarrow \mathcal{T}_n\\ \sigma_{i} &\mapsto \tau_{n-i}\end{cases}
\quad\mbox{ is an algebra isomorphism.}
$$
\end{prop}
 \prf The isomorphism is proved by direct computations.
\finprf

\subsection{Baxterisation}

The following theorem contains the main result of this paper and justifies the introduction of the algebra $\cS_n$.
 \begin{thm}\label{th:main}
   If $\sigma_i$ satisfy the relations of $\mathcal{S}_n$, then 
    \begin{eqnarray}
 \check R_i^\sigma(x,y)= \Sigma_i(y)  \Sigma_i(x)^{-1}\quad\text{where}\quad \Sigma_i(x)=1-x \sigma_i \label{eq:Smatrix-n}
 \end{eqnarray} 
 satisfy the braided Yang--Baxter equation
   \begin{equation}
   \check R_{i}(x,y) \check R_{i+1}(x,z)\check R_{i}(y,z)=\check R_{i+1}(y,z)\check R_{i}(x,z)\check R_{i+1}(x,y)\;. \label{eq-bybe}
   \end{equation}
Moreover the following properties hold: 
\begin{eqnarray}
&&\mb{-- unitarity }\
  \check  R_{i}(x,y)\check R_{i}(y,x)=1\;,\\[1.2ex]
&&\mb{-- regularity }
   \check R_{i}(x,x)=1\;,\\[1.2ex]
&&\mb{-- locality }\quad
   \check R_i(x,y)\check R_j(z,w)=\check R_j(z,w)\check R_i(x,y) \quad\text{for}\quad |i-j|>1\;.\label{eq:loca}
\end{eqnarray}
\end{thm}
\prf
The unitarity, regularity and locality properties are obvious.\\
To prove the braided Yang--Baxter equation \eqref{eq-bybe}, let us remark that, 
after multiplication  on the right by $\Sigma_i(y)$ and on the left by $\Sigma_{i+1}(y)$, it is equivalent to
\begin{eqnarray}
 A_i(y)A_i(x)^{-1}A_i(z)=A_i(z)A_i(x)^{-1}A_i(y)\label{eq:B}
\end{eqnarray}
where $A_i(x)=\Sigma_{i+1}(x)\Sigma_i(x)$.
Relation \eqref{eq:B} is equivalent to
\begin{equation}
 [\ A_i(x)\ ,\ A_i(y)\ ]=0\;. \label{eq:cB}
\end{equation}
Indeed, setting $z=0$ in \eqref{eq:B} leads to \eqref{eq:cB}, which obviously implies \eqref{eq:B}.
Let us notice now that
\begin{equation}
 A_i(x)=1-x(\sigma_i+\sigma_{i+1})+x^2 \sigma_{i+1}\sigma_i
\end{equation}
Therefore, the defining relation \eqref{eq-Sn-1} of $\cS_n$ implies relation \eqref{eq:cB} and then the braided Yang--Baxter equation.
\finprf
The parameters $x,y$ appearing for example in \eqref{eq:Smatrix-n} are called spectral parameters and a solution of the braided Yang--Baxter equation
is called a braided $R$-matrix.
Relation \eqref{eq:Smatrix-n} is called a Baxterisation: it is the construction of a braided $R$-matrix starting from an algebra with a finite number of generators.
Other Baxterisations appeared previously \cite{Jones,Jim86,Isaev} but starting from other algebras:
we want to emphasize that the Baxterisation introduced in the present paper depends 
\textsl{separately} on the two spectral parameters. This is a new feature in comparison to the previous 
Baxterisations.

Remark that the Baxterisation proposed here can be seen as a Drinfeld twist \cite{dri} of the permutation operator $P_{12}$ by the twist operator $F_{12}(z_1,z_2)=\Sigma_{1}(z_2)$.

Theorem \ref{th:main} gives a sufficient condition to obtain the braided Yang--Baxter equation. The following proposition proves that it is also a  necessary condition.
\begin{prop}\label{pro:1}
If $\check R_i^\sigma(x,y)$ given by \eqref{eq:Smatrix-n} satisfies the braided Yang--Baxter equation \eqref{eq-bybe} and the locality property \eqref{eq:loca}, then the generators
$\sigma_i$ satisfy $\cS_n$.
\end{prop}
\prf We have already seen that the braided Yang--Baxter equation implies \eqref{eq:cB}. The different coefficients of 
\eqref{eq:cB} w.r.t. $x$ and $y$ imply \eqref{eq-Sn-1}. The locality implies \eqref{eq-Sn-2} which concludes the proof.
\finprf

Until now, we used the algebra $\cS_n$ to get a solution of the braided Yang--Baxter equation, but we can use similarly the algebra $\cT_n$:
 \begin{thm}\label{th:main2}
The generators $\tau_i$ spend the algebra $\cT_n$ if and only if
    \begin{eqnarray}
 \check R_i^\tau(x,y)= \fT_i(x)  \fT_i(y)^{-1}\quad\text{where}\quad \fT_i(x)=1-x \tau_i \label{eq:Smatrix-np}
 \end{eqnarray} 
 are unitary, regular and local solutions of the braided Yang--Baxter equation.
\end{thm}
\prf
Direct consequence of theorem \ref{th:main} and proposition \ref{pro:1}, using the isomorphism of proposition \ref{pro:iso}.
\finprf
Let us stress that there is a flip between the spectral parameters in definitions \eqref{eq:Smatrix-n} and \eqref{eq:Smatrix-np}.

\subsection{Link with the Hecke algebra} 

The Hecke algebra $\cH_{n}(\xi)$ is the unital associative algebra over $\CC$ with generators $g_1,\dots,g_{n-1}$
and defining relations\footnote{Strictly speaking, the definition of the Hecke algebra states that the generators $g_i$ are invertible. We consider here, for later convenience, a more general version where this latter condition is relaxed.}
\begin{eqnarray}
&&g_ig_j=g_jg_i\quad \text{if}\quad |i-j|>1\\
&&g_ig_{i+1}g_i=g_{i+1}g_ig_{i+1}\\
&& g_i^2=g_i+\xi
\end{eqnarray}
where $\xi$ is a free parameter. There is a surjective homomorphism of algebra $\psi:\cS_n \rightarrow \cH_{n}(0)$ given on generators by 
 $\sigma_i \mapsto g_i$. Therefore, $\check R^H(z_1,z_2)=\psi(\Sigma(z_2)\,\Sigma(z_1)^{-1})$ satisfies the braided Yang--Baxter equation.
 It can be written as follows
 \begin{equation}
\check R^H_i(z_1,z_2)=1 + \frac{z_2-z_1}{z_1-1} g_i \;.
\end{equation}
With the change of variables $z_i=\lambda_i^2+1$, we recognize the usual Baxterisation of the Hecke algebra \cite{Jones}. We see that in this case, the $R$-matrix 
depends only on the ratio $\lambda_1/\lambda_2$.

\section{Representations of  $\cS_n$} 

In this section, we are interested in some matrix representations of $\cS_n$ and in particular the ones useful in the context of integrable systems.


\subsection{Representation in tensor product \label{sec:tens}}

We are interested in the matrix representations of $\cS_n$ in $End(\CC^m)^{\otimes n}$. 
More precisely, we look for representations of the following type
\begin{eqnarray}
 \cS_n &\rightarrow& End(\CC^m)^{\otimes n}\nonumber\\
 \sigma_i &\mapsto& \II ^{\otimes i-1} \otimes S \otimes \II ^{\otimes n-i-1}\label{eq:rep}
\end{eqnarray}
where $\II$ is the identity of $End(\CC^m)$ and $S\in End(\CC^m)\otimes End(\CC^m)$.
We use the following notation $S_{j,j+1}=\II^{\otimes j-1} \otimes S \otimes \II^{\otimes n-j-1}$:
the indices in $S_{j,j+1}$ label the copies of $End(\CC^m)$ in which the operator $S$ acts non-trivially.
To look for such matrix representations of $\cS_n$ ($n\geq 3$), it is necessary  and sufficient to find $S$ satisfying the single relation of $\cS_3$:
\begin{equation} \label{eq}
  [S_{23}\,S_{12}\,,\,S_{12}+S_{23}]=0\;.
\end{equation}
We give a classification of the solutions of this equation for $m=2$ in section \ref{sec:44} and some solutions for any $m$ in section \ref{sec:mm}. 

Before that, let us remark that  proposition \ref{pro:iso2} implies that if $S$ is a solution of \eqref{eq}, then so is $\fm_{\alpha,\beta,\gamma}(S)$.
We have loosely used the same notation for the M\"obius map acting on the algebra and the ones acting on the representation.
Moreover, in this matrix representation, new transformations appear which simplify the classification.
The useful ones are presented in the following lemma.
\begin{lem}\label{lem:sym}
If $S$ obeys relation \eqref{eq},
then 
\begin{itemize}
\item $S^{-1}$, when $S$ is invertible, 
\item $S_{21}^{t_1t_2}$, where $(.)^{t_1t_2}$ is the transposition in the space $End(\CC^m)\otimes End(\CC^m)$,
\item $Q_1Q_2\,S\,Q_2^{-1}Q_1^{-1}$, where $Q$ is any invertible element of $End(\CC^m)$,
\end{itemize}
all obey \eqref{eq}.
\end{lem}
\prf For each of the three cases, simple algebraic manipulations lead to the result. \finprf

Using the Baxterisation introduced in theorem \ref{th:main} and the realisation of $\cS_n$ given by \eqref{eq:rep}, the matrix
\begin{equation}
 \check R(x,y)=(\II-yS)(\II-xS)^{-1}\;\label{eq:Rc2}
\end{equation}
is a solution of the braided Yang--Baxter equation
  \begin{equation}
   \check R_{i,i+1}(x,y) \check R_{i+1,i+2}(x,z)\check R_{i,i+1}(y,z)=\check R_{i+1,i+2}(y,z)\check R_{i,i+1}(x,z)\check R_{i+1,i+2}(x,y)\;, \label{eq-bybe2}
   \end{equation}
 where the indices stand for the spaces on which the matrix $\check R(x,y)$ acts non-trivially.
 
Remark that the transformations on $S$ presented in the lemma \ref{lem:sym} induce the following transformations on $\check R(z_1,z_2)$ which preserves its properties 
listed in theorem \ref{th:main} :
\begin{itemize}
\item $\check R_{12}(z_1,z_2)\ \mapsto\ \check R_{12}(\frac1{z_1},\frac1{z_2})$, which can be generalized to $\check R_{12}(z_1,z_2)\ \mapsto\ \check R_{12}(\lambda(z_1),\lambda(z_2))$, for any function $\lambda$.
 \item $\check R_{12}(z_1,z_2)\ \mapsto\ \check R_{21}(z_1,z_2)^{t_1t_2}$,
 \item $\check R(z_1,z_2)\ \mapsto\ Q_1Q_2\check R(z_1,z_2)Q_2^{-1}Q_1^{-1}$.
\end{itemize}
 
 Similarly, one gets matrix representations for $\cT_n$. Indeed, we look for representations of the following type
 \begin{eqnarray}
 \cT_n &\rightarrow& End(\CC^m)^{\otimes n}\nonumber\\
 \tau_i &\mapsto& \II^{\otimes i-1} \otimes T \otimes \II^{\otimes n-i-1}\label{eq:repT}
\end{eqnarray}
In this case, one has to solve the equation 
 \begin{equation} \label{eq2}
  [T_{12}\,T_{23}\,,\,T_{12}+T_{23}]=0\;
\end{equation}
the associated $R$-matrix being now 
\begin{equation}
 \check R(x,y)=(\II-xT)(\II-yT)^{-1}\;. \label{eq:Rc3}
\end{equation}

At the algebraic level $\cS_n$ and $\cT_n$ are isomorphic, see proposition \ref{pro:iso}. At the level of the representations presented here, there
are more connections presented in the following lemma.
\begin{lem}\label{lem:ST}\ 
\begin{itemize}
\item[(i)] $S$ is a solution of equation \eqref{eq} if and only if $T=S^{t_1t_2}$ is a solution of equation \eqref{eq2}.
\item[(ii)] $S$ is a solution of equation \eqref{eq} if and only if $T=P\,S\,P$ is a solution of equation \eqref{eq2} (where $P$ is the permutation matrix).
\end{itemize}
\end{lem}
\prf Again, the proofs are based on simple algebraic manipulations. \finprf

 \subsection{Applications for integrable systems \label{sec:int}}
 
The Yang--Baxter equation plays a crucial role in the context of quantum integrable systems since the $R$-matrices in this context lead 
easily to integrable Hamiltonians \cite{STF}. We briefly recall this construction in this section and
exhibit the form of the Hamiltonian obtained for the $R$-matrix introduced previously.

Starting from the braided $R$-matrices \eqref{eq:Rc2}, we define the (unbraided) $R$-matrices as
 $R_{i,i+1}(x,y) = \check R_{i,i+1}(x,y)P_{i,i+1}$ where $P_{i,i+1}$ 
is the permutation operator acting on the $i^\text{th}$ and $(i+1)^\text{th}$ spaces.
We introduce the following transfer matrix 
\begin{equation}
 t(x\,|\,z)=tr_0 R_{01}(x,z)\dots R_{0n}(x,z)\;.
\end{equation}
The variable $z$  is  a free parameter playing the role of an inhomogeneity parameter. 
The main property of the transfer matrix is that it commutes for different values of its spectral parameter \textit{i.e.} $[t(x\,|\,z),t(y\,|\,z)]=0$.
Therefore, the operator defined by
\begin{equation}
 H=\left.\frac{d}{dx}t(x\,|\,z)\right|_{x=z} \  t(z\,|\,z)^{-1}
 \end{equation}
is an integrable Hamiltonian.
   
Since the $R$-matrix is regular, the Hamiltonian can be written as follows
\begin{equation}
 H=\sum_{j=1}^n h_{j-1,j}  \qquad 
\end{equation}
where by convention $h_{01}\equiv h_{n1}$ and
  \bea \label{eq:h}
h &=& \frac{d}{dz_1}\check R(z_1,z_2)\Big|_{z_1=z_2=z} = -\frac{d}{dz_2}\check R(z_1,z_2)\Big|_{z_1=z_2=z} 
= S (\II-zS)^{-1}=\fm_{0,1,-z}(S)\;.
\eea

Proposition \ref{pro:iso2} ensures that we can work modulo the action of  M\"obius transformations, 
so that  we can take  $h= S$. 
Then, the whole set of integrable Hamiltonians linked to the construction presented here is obtained by classifying 
the solutions $S$ of relation \eqref{eq} and by considering them as Hamiltonians.  
Moreover, using the lemma \ref{lem:ST}, it can be shown that the transformations
\beq\label{transfo-ST}
h \mapsto Q \otimes Q\ h^{t_1t_2}\ (Q\otimes Q)^{-1} \mb{and} h \mapsto Q \otimes Q\ P\ h\,P\ (Q \otimes Q)^{-1}
\eeq
also preserve integrability, based on the $R$-matrix \eqref{eq:Rc3}.

\section{Classification of the $4 \times 4$ representations \label{sec:44}}

Relation \eqref{eq} is difficult to solve in general: there are $(m^3)^2$ cubic relations in terms of the $(m^2)^2$ entries of the matrix $S$.
However, for $m=2$, using the transformations of lemma \ref{lem:sym} and a direct resolution with a formal mathematical software, we are able to compute all the solutions 
which are presented in the following theorem. Note that we do not impose
 a priori that $S$ is invertible and indeed some particular cases of the solutions are not.
\begin{thm}\label{th:cla4}
The whole set of representations of $\cS_n$ of type \eqref{eq:rep} for $m=2$ is obtained by applying all the possible
transformations presented in lemma \ref{lem:sym} or M\"obius transformation to the seven following matrices:
\begin{itemize}
\item two $5$-parameter matrices
\begin{equation}
 S^{(1)}=\begin{pmatrix}
                a & \cdot & \cdot & \cdot \\
                b & c & \cdot & \cdot \\
                d & \cdot & a & \cdot \\
                \cdot & e & \cdot & a 
               \end{pmatrix}, \quad
 S^{(2)}=\begin{pmatrix}
                a & b & c & d \\
                \cdot & a & \cdot & e \\
                \cdot & \cdot & a & b+c-e \\
                \cdot & \cdot & \cdot & a
               \end{pmatrix}
\end{equation}
\item four $4$-parameter matrices
\begin{equation}
 S^{(3)}=\begin{pmatrix}
          a & \cdot & \cdot & \cdot \\
          b & a+ \frac{b(d-a)}{c} & \cdot & \cdot \\
          c & \cdot & d & \cdot \\
          \cdot & \cdot & \cdot & a
         \end{pmatrix}, \quad 
 S^{(4)}=\begin{pmatrix}
          a & \cdot & \cdot & \cdot \\
          \cdot & b & \cdot & \cdot \\
          \cdot & c & a & \cdot \\
          \cdot & \cdot & \cdot & d
         \end{pmatrix}, 
         \end{equation}
  \begin{equation}
 S^{(5)}=\begin{pmatrix}
          a & \cdot & \cdot & \cdot \\
          \cdot & b & \cdot & \cdot \\
          \cdot & \cdot & c & \cdot \\
          \cdot & \cdot & \cdot & d
         \end{pmatrix}, \quad 
S^{(6)}=\begin{pmatrix}
          a & \cdot & \cdot & \cdot \\
          b & c & \cdot & d \\
          \cdot & \cdot & a & \cdot \\
          \cdot & -b & b & a 
         \end{pmatrix}, \quad 
  \end{equation}
\item one $3$-parameter matrix
\begin{equation}
 S^{(7)}=\begin{pmatrix}
          a & \cdot & \cdot & \cdot \\
          b & c & c-a & \cdot \\
          \cdot & \cdot & a & \cdot \\
          \cdot & \cdot & \cdot & c 
         \end{pmatrix},
\end{equation}
\end{itemize}
where $a,b,c,d,e$ are free complex parameters.
\end{thm}
\prf
The proof consists in finding all the $4\times 4$ matrices $S$ solution of equation \eqref{eq}.
We use a technique introduced in \cite{Hiet}  to classify the constant $4\times 4$ 
solutions of the Yang--Baxter equation.
Let $s_{ij}$ (for $i,j=1,2,3,4$) be the entries of $S$.
Using the transformations exposed in lemma \ref{lem:sym}, we can always look for solution of the equation with $s_{41}=0$. Indeed if $S$ is a 
solution with $s_{41} \neq0$, then $S$ is related through a transformation to a solution $S^{new}$ with $s^{new}_{41}=0$. More precisely if $s_{14}=0$ we set
$S^{new}=(S_{21})^{t}$ and we have $s^{new}_{41}=0$ because the transformation exchanges $s_{41}$ and $s_{14}$. If $s_{14}s_{41}\neq 0$, we set
$S^{new}=Q\otimes Q\,S\,Q^{-1}\otimes Q^{-1}$, with 
\begin{equation}
 Q=\begin{pmatrix}
    1 & 0 \\
    B & 1
   \end{pmatrix}.
\end{equation}
We have $s^{new}_{41}=s_{14}B^4+(s_{24}+s_{34}-s_{12}-s_{13})B^3+(s_{44}-s_{22}-s_{23}-s_{32}-s_{33}+s_{11})B^2+(s_{21}+s_{31}-s_{42}-s_{43})B+s_{41}$.
Since we are in the case $s_{14} \neq 0$, it is always possible to find $B$ such that $s^{new}_{41}=0$.
Therefore, without loss of generality, we can now set $s_{41}=0$.

At this stage, we use a computer software to solve \eqref{eq} with $s_{41}=0$. Then, using the transformations of lemma \ref{lem:sym}, we select the solutions 
which are not related by transformations and get the seven solutions presented in the theorem. \finprf

Remark that the solution $S^{(4)}$ provides the TASEP Markovian matrix as a subcase.

As explained in section \ref{sec:int}, each $S$ is an integrable Hamiltonian. Then, from any $4\times4$ integrable Hamiltonian $h$ obtained in this classification, we can construct new ones by action of the group of transformations generated by
\bea\label{transfo-H}
&&h \mapsto Q \otimes Q\, h\, (Q\otimes Q)^{-1} \mb{;} h \mapsto \ h^{t_1t_2} \mb{;} h \mapsto \ P\,h\,P
 \mb{;} h \mapsto (\alpha + \beta h)(1+\gamma h)^{-1}.\qquad
\eea

Using theorem \ref{th:main}  and the previous classification theorem \ref{th:cla4}, we get a set of solutions 
to the braided Yang--Baxter 
equation. The braided $R$-matrices are easily computed using \eqref{eq:Smatrix-n} but to simplify the use of this article, we give their explicit form in appendix 
\ref{app:R}. One has to keep in mind that one can construct other $R$-matrices from these ones by using the transformations described in section \ref{sec:tens}.

\section{ $m \times m$ representations \label{sec:mm}}

As explained previously, the resolution of equation \eqref{eq} is complicated and a complete classification for any $m$ seems impossible.
However, it is still possible to find some solutions. In the following proposition, we present particular non-trivial matrix representations for any $m$:
\begin{prop}\label{prop:S}
Let $E_{ij}$ be the canonical basis of $End(\CC^m)$ and let $\rho_i$ and $\mu_{i,j}$, $1\leq i<j\leq m$ be some complex numbers.
We define
\begin{eqnarray}
 {S}&=&\sum_{1\leq i<j \leq m} \rho_i\ E_{ii}\otimes E_{jj} +\mu_{i,j}\ E_{ji}\otimes E_{ij}\label{eq:STASEP}
\end{eqnarray}
Then, the map $\sigma_i\mapsto {S}_{i,i+1}$ is an homomorphism of algebra $\mathcal{S}_n\rightarrow \big(End(\CC^m)\big)^{\otimes n}$. 
\end{prop}
\prf We prove by direct computations that $S$ given by \eqref{eq:STASEP} verifies \eqref{eq}.
\finprf
Remark that $S$ is non-invertible.
Obviously, the use of the transformations given in lemma \ref{lem:sym} provides new  representations isomorphic to \eqref{eq:STASEP}.

There is a similar type of representation for $\cT_n$. 
\begin{prop}\label{prop:T}
Let $E_{ij}$ be the canonical basis of $End(\CC^m)$ and let $\zeta_j$ and $\nu_{i,j}$, $1\leq i<j\leq m$ be some complex numbers.
We define
\begin{equation}
T=\sum_{1\leq i<j \leq m} \zeta_j\ E_{ii}\otimes E_{jj} +\nu_{i,j}\ E_{ji}\otimes E_{ij}\;,\label{eq:TTASEP}
\end{equation}
Then, the map $\tau_i\mapsto T_{i,i+1}$ is an homomorphism of algebra 
$\mathcal{T}_n\rightarrow \big(End(\CC^m)\big)^{\otimes n}$. 
\end{prop}
\prf 
Direct consequence of proposition \ref{prop:S} and transformations \eqref{transfo-ST} with 
Q=$\begin{pmatrix}  0 & 1\\  1 & 0   \end{pmatrix}$.
\finprf
Again, lemma \ref{lem:sym} provides isomorphic forms for the representations \eqref{eq:TTASEP}. Moreover, lemma \ref{lem:ST} provides new representations of $\cS_n$ (resp. $\cT_n$) starting from \eqref{eq:TTASEP} (resp. \eqref{eq:STASEP}).

From $S$ (resp. $T$) given by \eqref{eq:STASEP} (resp. \eqref{eq:TTASEP}), one gets an $R$-matrix using Baxterisation \eqref{eq:Rc2} (resp. \eqref{eq:Rc3}). Surprisingly enough, from these two $R$-matrices, we can obtain another one, as stated in the following theorem.

\begin{thm}\label{th:prod}
Let us define the matrix
\begin{equation}
 \check R(x,z)=\check R^\rho(x,z)\check R^\zeta(x,z)\; ,
\end{equation} 
where  $\check R^\rho(x,y)=(\II-yS)(\II-xS)^{-1}$ and  $\check R^\zeta(x,y)=(\II-xT)(\II-yT)^{-1}$
are respectively the Baxterisations of $S$ given by \eqref{eq:STASEP} and of $T$ given by \eqref{eq:TTASEP}.

If the relations $\rho_i\nu_{i,j}=\mu_{i,j}\zeta_j$ for $1\leq i<j\leq m$ hold, $\check R(x,z)$
satisfies the braided Yang--Baxter equation and is unitary.
\end{thm}
\prf It is proven by direct computations.\finprf
Let us remark that the relations $\rho_i\nu_{i,j}=\mu_{i,j}\zeta_j$ are equivalent to $[T,S]=0$.

The $R$-matrix introduced in the theorem is a generalisation of the matrix introduced in \cite{cantini} to study the multi-species totally 
asymmetric exclusion process with different hopping rates.

\section{Conclusion}

Numerous  questions are still open. 
The Hecke algebra has been a very useful tool in different contexts: \textit{e.g.} it is the centralizer of the quantum group $\cU_q(gl_N)$ \cite{Jim86} 
or it permits to construct link invariants \cite{jones2}. We believe that the algebra $\cS_n$ we have introduced here should have similar fields of applications.
We think that the classification of its irreducible representations should be also interesting.
The defining relations of the algebra $\cS_n$ look also very similar to the ones of the braid group: the connections between these two algebras should also be explored.

In theorem \ref{th:prod}, we show that the product
of two $R$-matrices based on $\cS_n$ and $\cT_n$ provides a new $R$-matrix if a simple condition on the parameters holds. 
It would be interesting to understand if this feature is 
associated to the special representation used in the theorem or if it is still true at the algebraic level.

Finally, the list of $R$-matrices provided in this paper may be 
used to introduce new models in the context of quantum mechanics (spin chains), 2D-statistical models (loop or vertex models) 
or Markovian models (exclusion processes). The knowledge of their associated $R$-matrix may allow one to solve them using, for example, 
the algebraic Bethe ansatz \cite{STF} or the matrix ansatz (see \cite{DEHP,BE} for an introduction and \cite{SW,CRV} for its use starting from
 $R$-matrices).

\appendix
\section{$R$-matrices from the classification theorem \ref{th:cla4} \label{app:R}}

The braided R-matrices associated to the $S$'s of theorem \ref{th:cla4} are respectively: 
\begin{equation}
  \check R^{(1)}(x,y)=
 \begin {pmatrix} {\frac {ya-1}{xa-1}}&0&0&0
\\ \noalign{\medskip}{\frac {b \left( x-y \right) }{ \left( xc-1
 \right)  \left( xa-1 \right) }}&{\frac {yc-1}{xc-1}}&0&0
\\ \noalign{\medskip}{\frac {d \left( x-y \right) }{ \left( xa-1
 \right) ^{2}}}&0&{\frac {ya-1}{xa-1}}&0\\ \noalign{\medskip}-{\frac {
exb \left( x-y \right) }{ \left( xc-1 \right)  \left( xa-1 \right) ^{2
}}}&{\frac {e \left( x-y \right) }{ \left( xc-1 \right)  \left( xa-1
 \right) }}&0&{\frac {ya-1}{xa-1}}\end {pmatrix}
\end{equation}

\begin{equation}
\check R^{(2)}(x,y)=
 \begin{pmatrix} {\frac {ya-1}{xa-1}}&{\frac {b \left( x-y
 \right) }{ \left( xa-1 \right) ^{2}}}&{\frac {c \left( x-y \right) }{
 \left( xa-1 \right) ^{2}}}&{\frac { \left( x-y \right)  \left( adx-xc
b-exb-x{c}^{2}+xce-d \right) }{ \left( xa-1 \right) ^{3}}}
\\ \noalign{\medskip}0&{\frac {ya-1}{xa-1}}&0&{\frac {e \left( x-y
 \right) }{ \left( xa-1 \right) ^{2}}}\\ \noalign{\medskip}0&0&{\frac 
{ya-1}{xa-1}}&{\frac { \left( b+c-e \right)  \left( x-y \right) }{
 \left( xa-1 \right) ^{2}}}\\ \noalign{\medskip}0&0&0&{\frac {ya-1}{xa
-1}}\end{pmatrix} 
\end{equation}

\begin{equation}
\check R^{(3)}(x,y)=  \begin{pmatrix} {\frac {ya-1}{xa-1}}&0&0&0
\\ \noalign{\medskip}-{\frac {bc \left( x-y \right) }{ \left( xa-1
 \right)  \left( abx-acx-bdx+c \right) }}&{\frac {aby-acy-bdy+c}{abx-a
cx-bdx+c}}&0&0\\ \noalign{\medskip}{\frac {c \left( x-y \right) }{
 \left( xa-1 \right)  \left( xd-1 \right) }}&0&{\frac {yd-1}{xd-1}}&0
\\ \noalign{\medskip}0&0&0&{\frac {ya-1}{xa-1}}\end{pmatrix}
\end{equation}

\begin{equation}
 \check R^{(4)}(x,y)=
  \begin{pmatrix} {\frac {ya-1}{xa-1}}&0&0&0
\\ \noalign{\medskip}0&{\frac {yb-1}{xb-1}}&0&0\\ \noalign{\medskip}0&
{\frac {c \left( x-y \right) }{ \left( xb-1 \right)  \left( xa-1
 \right) }}&{\frac {ya-1}{xa-1}}&0\\ \noalign{\medskip}0&0&0&{\frac {y
d-1}{xd-1}}  \end{pmatrix}
\end{equation}

\begin{equation}
 \check R^{(5)}(x,y)=
  \begin{pmatrix}{\frac {ya-1}{xa-1}}&0&0&0
\\ \noalign{\medskip}0&{\frac {yb-1}{xb-1}}&0&0\\ \noalign{\medskip}0&0
&{\frac {yc-1}{xc-1}}&0\\ \noalign{\medskip}0&0&0&{\frac {yd-1}{xd-1}}
 \end{pmatrix}
\end{equation}

\begin{eqnarray}
 \check R^{(6)}(x,y)&=&  \begin{pmatrix} {\frac {ya-1}{xa-1}}&0&0&0\\ 
 \noalign{\medskip}{\frac {b \left( x-y \right) }{\fn(x,x)}}
 &{\frac {\fn(x,y)}{\fn(x,x)}}
 &-{\frac {bdx \left( x-y \right) }{ \left( xa-1 \right) \fn(x,x) }}
 &{\frac {d \left( x-y \right) }{\fn(x,x)}}\\ 
  \noalign{\medskip}0&0&{\frac {ya-1}{xa-1}}&0\\ 
 \noalign{\medskip}{\frac {{b}^{2}x \left( x-y \right) }{ \left( xa-1 \right)  \fn(x,x) }}
&-{\frac {b \left( x-y \right) }{\fn(x,x)}}
&{\frac {b \left( x-y \right)  \left( xc-1 \right) }{ \left( xa-1 \right)  \fn(x,x) }}
&{\frac {\fn(x,y)}{\fn(x,x)}} \end{pmatrix}
\\
\fn(x,y) &=& (ac+bd)xy-ax-cy+1
\end{eqnarray}

\begin{equation}
 \check R^{(7)}(x,y)=\begin{pmatrix}{\frac {ya-1}{xa-1}}&0&0&0
\\ \noalign{\medskip}{\frac {b \left( x-y \right) }{ \left( xc-1
 \right)  \left( xa-1 \right) }}&{\frac {yc-1}{xc-1}}&{\frac {
 \left( c-a \right)  \left( x-y \right) }{ \left( xc-1 \right) 
 \left( xa-1 \right) }}&0\\ \noalign{\medskip}0&0&{\frac {ya-1}{xa-1}}
&0\\ \noalign{\medskip}0&0&0&{\frac {yc-1}{xc-1}} \end{pmatrix}
\end{equation}


\end{document}